\documentclass[aps,prd,twocolumn,superscriptaddress,groupedaddress,amsmath,amssymb,longbibliography,nofootinbib]{revtex4-2}
\usepackage{amsmath,amsthm,amssymb}
\usepackage{graphicx}
\usepackage{hyperref}
\usepackage{xcolor}
\usepackage{booktabs}
\usepackage{tikz}
\usepackage{enumitem}
\usepackage{array}
\usepackage{subcaption}
\usepackage{afterpage}

\hypersetup{colorlinks=true,linkcolor=blue,citecolor=blue,urlcolor=blue}

\begin{document}

\title{Emergent Grand Unified Structure in a\\ $4 \times 4$ Nilpotent Matrix Algebra}

\author{M. Adeel Ajaib}
\affiliation{Pennsylvania State University, Abington College, Abington, PA 19001, USA}

\begin{abstract}

We show that nilpotent matrices that yield the Schrödinger equation from its first order form encode the fingerprints 
of grand unified theories. We perform a rigorous search for all such nilpotent matrices and find that the resulting matrices naturally organize into suggestive group theoretic structures without any other a priori assumptions. The antisymmetric sector consists of three groups of
sixteen matrices, each of which further splits as $16 = 12 + 4$ and  exhibits unique characteristics in the step
potential scattering problem. The symmetric zero-diagonal sector also forms three families, mirroring the quark--lepton
decomposition of the Pati-Salam model. 
These results may help answer why there are three families of fermions and also demonstrate that the $4\times4$ matrix
algebra is a compact, nontrivial shadow of the SO(10) embedding, with fermion-like
and gauge-like subspaces.
\end{abstract}

\maketitle

\section{Introduction}
\label{sec:intro}

The Standard Model contains three generations of quarks and leptons with an unexplained multiplicity structure. Grand Unified Theories, particularly SO(10)~\cite{Georgi1974,Fritzsch1974,Georgi1975}, provide elegant organization by embedding one generation into a single 16-dimensional spinor representation. Under the Pati-Salam subgroup~\cite{Pati1974}, this decomposes as:
\begin{equation}
\text{SO}(10) \supset \text{SU}(4)_C \times \text{SU}(2)_L \times \text{SU}(2)_R,
\end{equation}
with the spinor branching:
\begin{equation}
\mathbf{16} \to (4,2,1) \oplus (\bar{4},1,2).
\end{equation}

Further decomposition under SU(4)$_C \to$ SU(3)$_C \times$ U(1)$_{B-L}$ yields:
\begin{equation}
\mathbf{16} \to 3(3,2) \oplus 3(\bar{3},1) \oplus (1,2) \oplus (1,1),
\end{equation}
containing exactly 12 quarks and 4 leptons.

Here, we demonstrate that a discrete 4$\times$4 nilpotent matrix algebra, emergent from 1+1D potential-step scattering~\cite{Ajaib2015, Ajaib2015b}, reproduces this exactly: 48 antisymmetric matrices as three left-chiral 16-plets, 48 symmetric zero-diagonal as three right-chiral 16-bars (with 12+4 Pati--Salam texture), and 384 symmetric into SO(10)-like gauge sectors (32+32+64+64+192). No continuous groups or extra dimensions are assumed---patterns arise from nilpotency and anti-commutation constraints.

Nilpotent factorizations of Dirac-type operators have appeared in several
algebraic approaches to particle physics
\cite{Rowlands1998,Rowlands2003,Diaz2007,Rowlands2018}.
Our construction differs in that we impose nilpotency on a restricted
$4\times 4$ matrix algebra tailored to nonrelativistic scattering and examine
the emergent SO(10)-like organization.

The paper is organized as follows. Section II details the construction of all 4$\times$4 nilpotent matrices satisfying the required constraints and demonstrates their organization into three families of 16-plets with universal (12+4) internal structure. Section III examines the quantum scattering signatures that distinguish the three generations through energy-dependent spin-flip transmission coefficients. Section IV establishes the correspondence with SO(10) grand unification, showing how the discrete matrix algebra reproduces both the spinor representation structure and the Pati-Salam decomposition. Section V presents experimentally testable predictions through spin-polarized electron scattering. We conclude in Section VI.

\section{Matrix Construction and Multiplicity Structure}
\label{sec:matrices}


Consider the quantum scattering formalism in~\cite{Ajaib2015, Ajaib2015b} where 4×4 matrices $\eta$ generate dynamics through the following equation:
\begin{equation}
p_z = \eta E + \eta^\dagger m.
\label{eq:NR-1D}
\end{equation}

For this to yield consistent quantum mechanics, $\eta$ must satisfy:
\begin{subequations}\label{eq:constraints}
\begin{align}
\eta^2 &= 0, \label{eq:nil1}\\
\{\eta, \eta^\dagger\} &= c\,I, \label{eq:norm}\\
\eta^T &= \pm \eta. \label{eq:trans}
\end{align}
\end{subequations}

Constraint (\ref{eq:nil1}) and (\ref{eq:norm}) ensure proper dispersion relations and normalization. Constraint (\ref{eq:trans}) distinguishes symmetric and antisymmetric cases.
We restrict matrix entries to the discrete set \{0, ±1, ±i\}. Our search yields that the space of matrices can be organized into symmetric and antisymmetric subsets with nontrivial internal structure:

\begin{itemize}
\item A set of 48 nonzero antisymmetric matrices with vanishing diagonals, which naturally arrange into three families of 16 with identical scattering behaviour. Each family of 16 matrices further decomposes as 16 = 12 + 4 according to algebraic criteria on their off-diagonal blocks:
\begin{equation}
48_{\text{antisym}} = 16_1 + 16_2 + 16_3,
\end{equation}

\item A set of 48 symmetric zero-diagonal matrices, which also split into three 16-plets depicting similar traits to the antisymmetric matrices:

\begin{equation}
48_{\text{sym}} = 16'_1 + 16'_2 + 16'_3.
\end{equation}

\item Additional symmetric matrices grouped into highly structured subsets of sizes 32, 32, 64, 64, and 192, characterized by their 2$\times$2 diagonal blocks in the Pauli basis.
\end{itemize}

Note that the search for asymmetric nilpotent matrices is computationally more expensive, and these matrices do not exhibit the physically relevant spin or transmission patterns seen in the symmetric and antisymmetric sectors. So in this study we focus on the above cases only.




Every admissible matrix has 2×2 block form:
\begin{equation}
M = \begin{pmatrix} A & X \\ Y & B \end{pmatrix},
\end{equation}
where A, B are diagonal blocks and X, Y are off-diagonal blocks.

The (12+4) Pati-Salam decomposition emerges universally across all three generations through a concrete mathematical criterion: the trace of the upper off-diagonal $2\times2$ block. For each matrix $M$, computing $\text{Tr}(X)$ systematically separates the 16-element families into 12-element and 4-element subsets. In Generation 1 (zero diagonal blocks), the division occurs between matrices with real trace ($\text{Tr}(X) \in \{0, \pm2\}$, 12 matrices) and imaginary trace ($\text{Tr}(X) \in \{\pm2i\}$, 4 matrices). In Generations 2 and 3 (non-zero diagonal blocks), the same (12+4) structure appears through traceless blocks ($\text{Tr}(X) = 0$, 12 matrices) versus traceful blocks ($\text{Tr}(X) \neq 0$, 4 matrices).

This demonstrates that the Pati-Salam decomposition is not an observation but a computable algebraic property present in all three generations, with the criterion adapting systematically to the diagonal structure.

\section{Quantum Dynamics and Scattering}
\label{sec:scattering}

\subsection{Dispersion Relations}

In this Section, we review how the non-relativistic and relativistic dispersion relations are obtained from the nilpotent matrices. We focus on the relativistic case and evaluate the transmission coefficients for all the matrices found in our analysis discussed in Section \ref{sec:matrices}. Consider equation (\ref{eq:NR-1D}):
\begin{equation}
p_z = \eta E + m\eta^\dagger.
\end{equation}
Squaring the momentum operator yields:
\begin{equation}
p_z^2 = E^2 \eta^2 + m^2 (\eta^\dagger)^2 + Em\{\eta,\eta^\dagger\}.
\end{equation}

Using the constraints $\eta^2 = 0$, $(\eta^\dagger)^2 = 0$, and $\{\eta,\eta^\dagger\}=2I$:
\begin{equation}
p_z^2 = 2Em,
\end{equation}
which gives the nonrelativistic dispersion relation $E = p_z^2/2m$.

For relativistic extension, we have \cite{Ajaib:2025guv}:
\begin{equation}
p_z = \chi_1 E - m\chi_2.
\end{equation}
where,
\begin{equation}
\chi_1 = \frac{\eta + \eta^\dagger}{\sqrt{2}}, \quad \chi_2 = \frac{\eta - \eta^\dagger}{\sqrt{2}},
\end{equation}
satisfying $\chi_1^2 = I$, $\chi_2^2 = -I$, $\{\chi_1,\chi_2\}=0$. These yield:
\begin{equation}
E^2 = p_z^2 + m^2.
\end{equation}

Both Schrödinger and Dirac dispersion relations emerge from the nilpotent algebra. The matrices $\chi_1$ and $\chi_2$ are evaluated for each of the $\eta$ matrices found in our search and we observe that the transmission coefficients exhibit 3 unique patterns for the 48 zero block diagonal anti-symmetric/symmetric cases.

\begin{figure*}[p]
\centering
\begin{subfigure}[b]{0.45\textwidth}
    \centering
    \includegraphics[width=\textwidth]{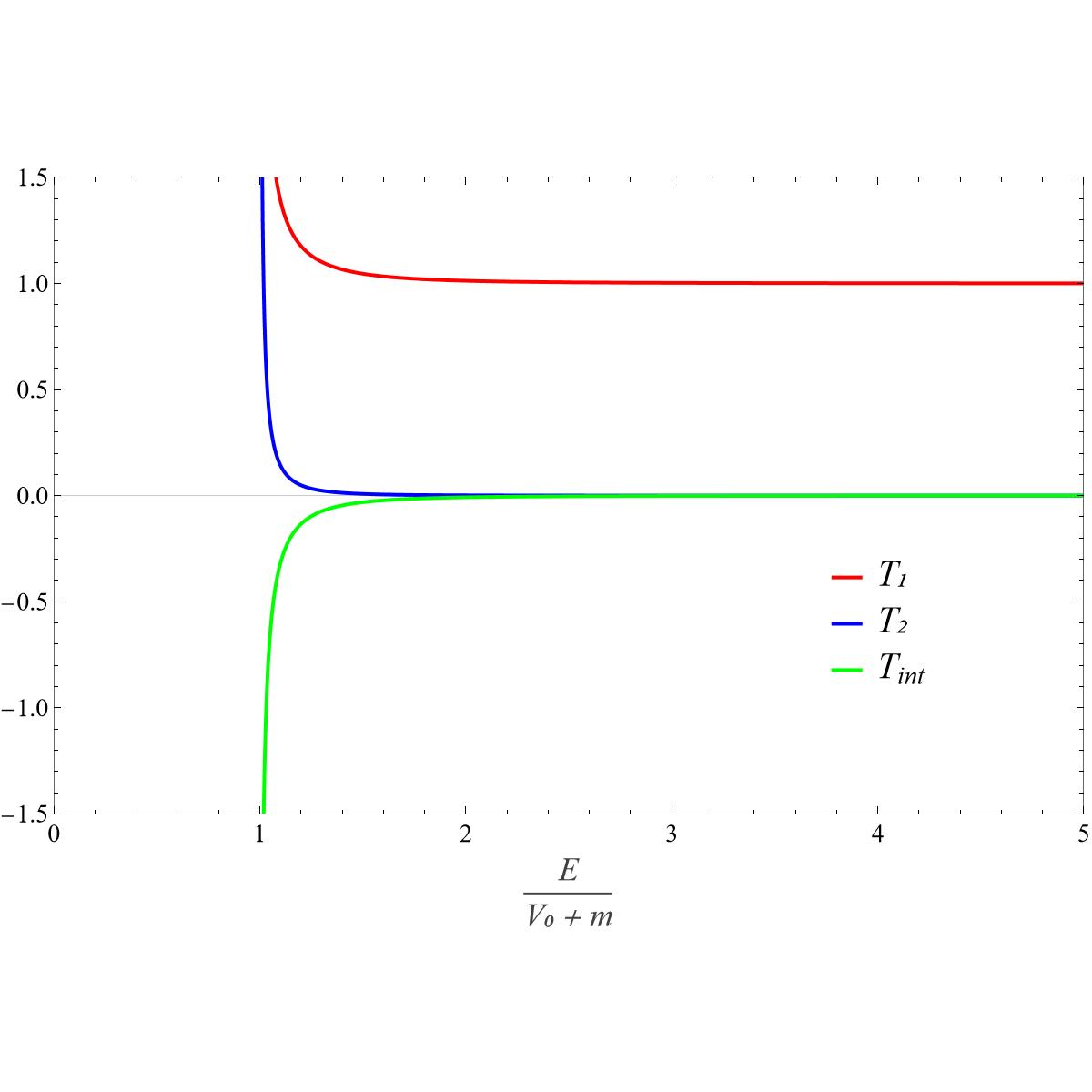}
    \caption{Generation 1}
    \label{fig:gen1}
\end{subfigure}
\hfill
\begin{subfigure}[b]{0.45\textwidth}
    \centering
    \includegraphics[width=\textwidth]{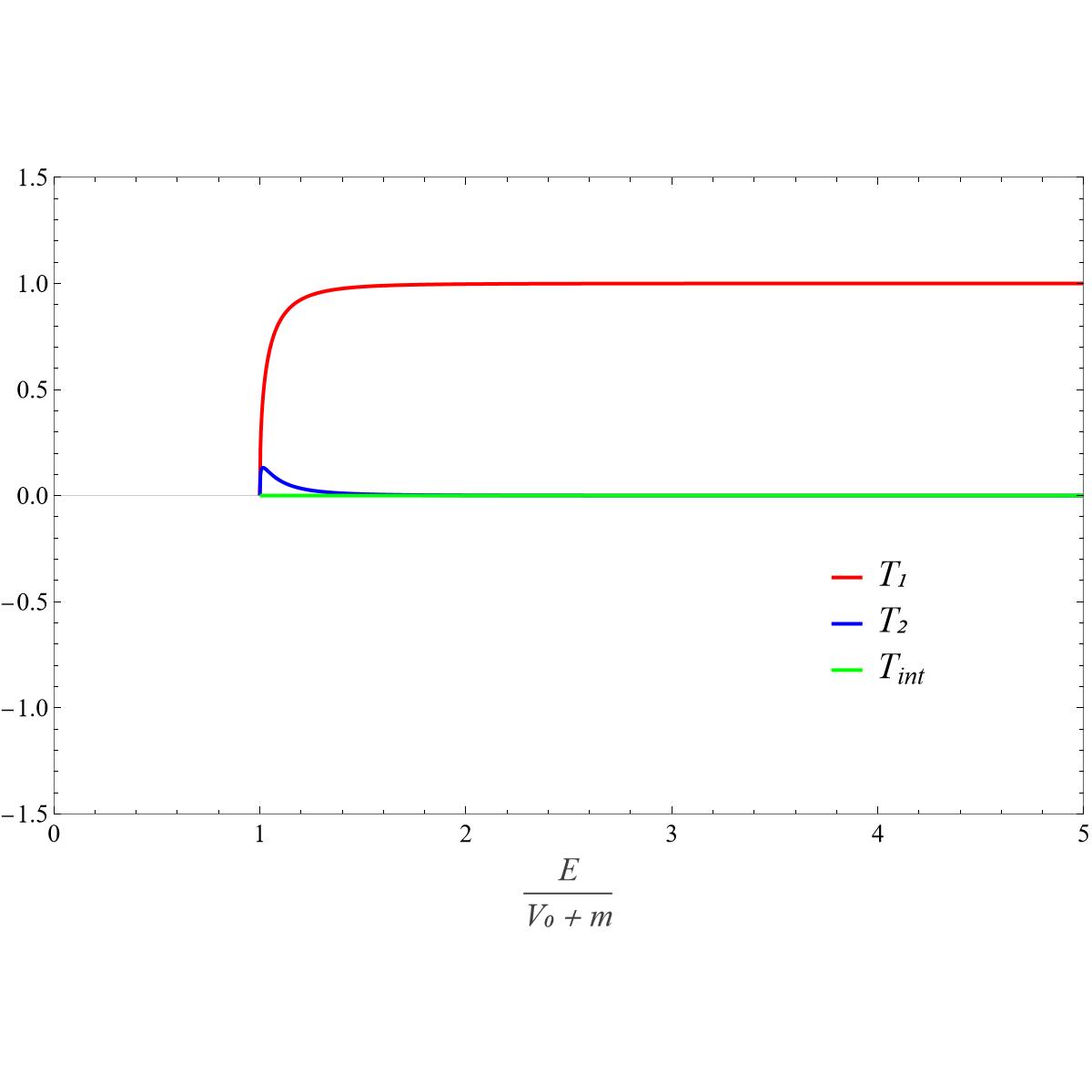}
    \caption{Generation 2}
    \label{fig:gen2}
\end{subfigure}

\vspace{0.5cm}

\begin{subfigure}[b]{0.45\textwidth}
    \centering
    \includegraphics[width=\textwidth]{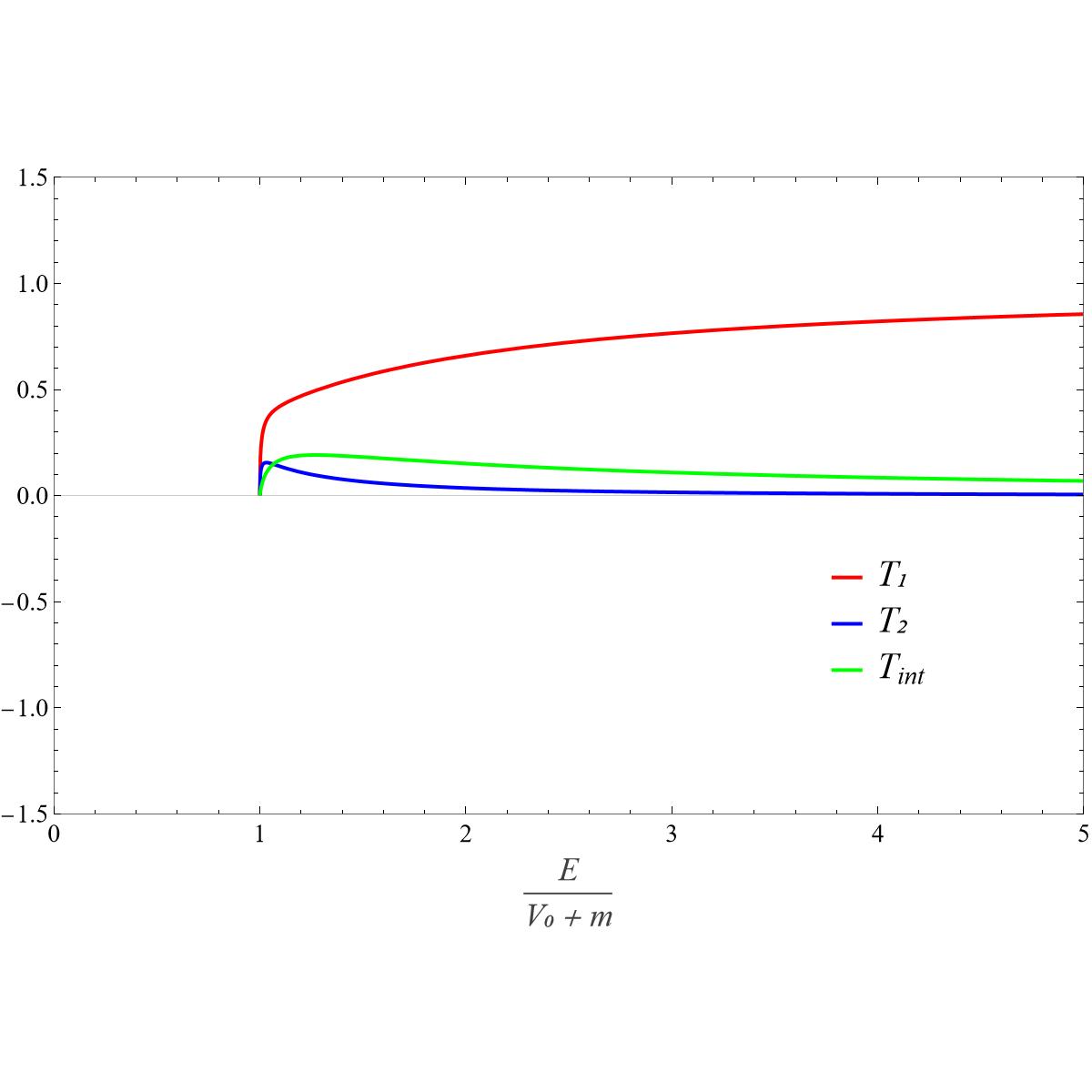}
    \caption{Generation 3}
    \label{fig:gen3}
\end{subfigure}
\caption{Relativistic transmission coefficients for spin-up electrons incident on a step potential for the three antisymmetric matrix families as functions of energy $E/(V_0+m)$. $T_1$ (red): transmitted as spin-up (no flip). $T_2$ (blue): transmitted as spin-down (spin-flip). $T_{\text{int}}$ (green): interference term. (a) Generation 1 (zero diagonal blocks) shows sharp threshold with strong spin-flip activity near threshold, characteristic of light particles. (b) Generation 2 (real diagonal entries) displays intermediate behavior with suppressed spin-flip. (c) Generation 3 (complex diagonal entries) exhibits smooth threshold with most suppressed spin-flip, characteristic of heavy particles. }
\label{fig:three_generations}
\end{figure*}

\subsection{Transmission Patterns and Generation Structure}

The three antisymmetric 16-plets exhibit fundamentally distinct scattering signatures, as shown in Fig.~\ref{fig:three_generations}. We consider spin-up electrons approaching a step potential. Each family produces unique energy-dependent transmission coefficients: $T_1(E)$ (transmitted as spin-up, no spin-flip), $T_2(E)$ (transmitted as spin-down, spin-flip), and $T_{\text{int}}(E)$ (spin interference term) in relativistic step-potential scattering. The explicit analytical 
expressions for the generation-dependent transmission coefficients are 
given in Appendix~\ref{app:transmission}.

\textbf{Generation 1} [Fig.~\ref{fig:gen1}]: Matrices have \emph{zero diagonal blocks} (A = B = 0), characteristic of massless or nearly massless particles. Shows the sharpest threshold behavior at $E \approx V_0 + m$ with \emph{strong} spin-flip activity near threshold. The significant spin-flip channel $T_2$ indicates strong chirality mixing. The system rapidly transitions to predominantly spin-preserving transmission ($T_1 \to 1$) at higher energies. 

\textbf{Generation 2} [Fig.~\ref{fig:gen2}]: Matrices have \emph{real non-zero diagonal entries} from \{0, ±1\}, display intermediate threshold behavior with suppressed spin-flip transitions compared to Generation 1. The spin-flip channel $T_2$ shows reduced amplitude throughout the energy range. The spin-preserving channel $T_1$ reaches partial transmission ($T_1 \approx 0.8$ at high energy), and the interference term $T_{\text{int}}$ remains small.

\textbf{Generation 3} [Fig.~\ref{fig:gen3}]: Matrices have \emph{complex diagonal entries} from \{0, ±1, ±i\}, indicating the richest diagonal structure. Exhibits the smoothest threshold with \emph{most suppressed} spin-flip transitions relative to Generation 1. While spin-flip activity is present near threshold (as in all three generations), it remains suppressed. The gradual onset of transmission and smooth threshold behavior are characteristic of heavy particles with weak chirality mixing. 

All three generations exhibit spin-flip transmission ($T_2 \neq 0$) near threshold, demonstrating that spin-mixing occurs in all cases. However, the spin-flip strength varies systematically: Generation 1 (zero diagonal) shows the \emph{strongest} spin-flip, characteristic of light/massless particles with maximal chirality mixing. Generations 2 and 3 show progressively \emph{suppressed} spin-flip as the diagonal structure becomes more complex, consistent with increasing mass and decreasing relativistic effects.

\section{SO(10) Correspondence}

The 96 matrices exhibit a structural correspondence with SO(10) grand unification across multiple independent features. This section establishes the correspondence systematically, progressing from numerical counting to detailed structure to physical interpretation. Each correspondence is exact and emerges from the same algebraic constraints. The probability of six independent numerical coincidences appears small, suggesting deeper structure rather than accident.

\subsection{Pati-Salam Decomposition: (12+4) Structure}

The Pati-Salam subgroup $SU(4)_C \times SU(2)_L \times SU(2)_R$ decomposes one SO(10) generation (16-dimensional spinor) into quarks and leptons with precise multiplicities:
\begin{equation}
\mathbf{16} \to (4,2,1) \oplus (\bar{4},1,2)
\end{equation}

This particle content decomposes naturally by chirality: the 12 quark states consist of 6 left-handed quarks (3 colors $\times$ 2 weak doublet components) and 6 right-handed quarks (3 colors $\times$ 2 components for up-type and down-type), while the 4 lepton states comprise 2 left-handed leptons forming a weak doublet ($\nu$, $e^-$) and 2 right-handed leptons appearing as singlets ($e^-_R$, $\nu_R$). This $(6_L + 6_R) + (2_L + 2_R) = 8_L + 8_R$ organization per generation precisely matches the left-right symmetric structure of Pati-Salam models, where the $SU(2)_L \times SU(2)_R$ gauge symmetry treats left- and right-handed fermions on equal footing, predicting the existence of right-handed neutrinos and providing a natural framework for understanding parity violation as a consequence of spontaneous symmetry breaking rather than a fundamental asymmetry of nature.

The 16-matrix structure of each generation exhibits a remarkable left-right symmetric organization: $16 = 8_L + 8_R$, precisely matching the Pati-Salam $SU(2)_L \times SU(2)_R$ decomposition. The 12 quark matrices organize as six left-right doublets, while the 4 lepton matrices split as two left-handed and two right-handed states. This structure manifests through two distinct $Z_2$ symmetries that act on different matrix blocks. The $\sigma_1$ symmetry, $(X,Y) \to (-X,-Y)$, operates universally across all three generations, creating six doublet pairings in the quark sector. For Generation 1 (zero diagonal blocks), $\sigma_1$ is the sole active symmetry relating states.

\subsection{Emergent Chiral Structure}

The nilpotency constraints combined with transpose symmetry force diagonal blocks into restricted subspaces that differ between sectors.

For the antisymmetric sector ($\eta^T = -\eta$), the transpose constraint forces diagonal blocks to be antisymmetric: $A^T = -A$, $B^T = -B$. Among the Pauli matrices, only $\sigma_y$ is antisymmetric ($\sigma_y^T = -\sigma_y$), yielding:
\begin{equation}
A, B \in \text{span}_{\mathbb{R}}\{\sigma_y\}.
\end{equation}

For the symmetric sector ($\eta^T = +\eta$), the transpose constraint forces diagonal blocks to be symmetric: $A^T = +A$, $B^T = +B$. While both $\sigma_x$ and $\sigma_z$ are symmetric, the nilpotency constraint combined with the discrete alphabet selects only:
\begin{equation}
A, B \in \text{span}_{\mathbb{R}}\{\sigma_x\}.
\end{equation}

Interestingly, $\sigma_z$ and the identity $I$ are systematically excluded from the diagonal blocks of the 48 symmetric zero-diagonal 16-plets, but present in the larger 384-element symmetric nilpotent sector. This exclusion is not imposed but emerges from the interplay of nilpotency ($\eta^2=0$), transpose symmetry ($\eta^T = \pm\eta$), and the discrete alphabet $\{0, \pm1, \pm i\}$.

The different Pauli structures between sectors provide a natural basis for interpreting them as conjugate representations:
\begin{align}
\text{Antisymmetric } (\sigma_y) &\leftrightarrow \mathbf{16} \sim (4,2,1) \nonumber\\
\text{Symmetric } (\sigma_x) &\leftrightarrow \overline{\mathbf{16}} \sim (\bar{4},1,2)
\end{align}

The first contains left-handed quarks and leptons, the second their right-handed counterparts. Both exhibit identical internal structure (3 families of 16 with (12+4) decomposition), differing only in the Pauli matrix appearing in diagonal blocks and chirality assignment—precisely as in SO(10).

\subsection{Lie Algebra Structure and Nilpotent Constraints}

The relationship between the 48 antisymmetric matrices and gauge algebra structure requires careful mathematical distinction. Let $\mathcal{A}$ denote the complex span of the 48 antisymmetric matrices:
\begin{equation}
\mathcal{A} = \text{span}_{\mathbb{C}}\{\eta_1, \eta_2, \ldots, \eta_{48}\}.
\end{equation}

We have verified computationally that $\dim_{\mathbb{C}}(\mathcal{A}) = 6$. Since commutators of antisymmetric matrices remain antisymmetric—$[A,B]^T = -[A,B]$ for all antisymmetric $A,B$—the span is closed under Lie brackets and forms a Lie algebra isomorphic to:
\begin{equation}
\mathcal{A} \cong \mathfrak{so}(4,\mathbb{C}) = \mathfrak{su}(2)_L \oplus \mathfrak{su}(2)_R.
\end{equation}

This establishes that the 48 matrices \emph{generate} the full chiral gauge algebra of the Pati-Salam subgroup.

However, the 48 matrices themselves are not arbitrary elements of $\mathfrak{so}(4,\mathbb{C})$—they satisfy the additional nilpotency constraint $\eta^2 = 0$. Nilpotent elements form an algebraic variety within the Lie algebra, not a Lie subalgebra: commutators of nilpotent matrices are generally not nilpotent. Indeed, direct computation shows $[\eta_i,\eta_j]^2 \neq 0$ for many pairs, confirming that while $[\eta_i,\eta_j] \in \mathcal{A}$ (Lie algebra closure), we have $[\eta_i,\eta_j] \notin \{\eta_1,\ldots,\eta_{48}\}$ (not closed as a discrete set).

The key observation is that the nilpotency constraint selects exactly 48 special elements within the 6-dimensional gauge algebra, and these 48 elements organize into three families of 16 with universal (12+4) decomposition. This demonstrates that \emph{representation structure emerges from algebraic constraints within the gauge algebra}, not from representation theory itself. The constraint $\eta^2=0$ naturally produces the triplication and internal Pati-Salam structure without imposing gauge invariance or group representations.

The symmetric sector exhibits complementary behavior: the 48 symmetric matrices also span a 6-dimensional space, but their span is \emph{not} closed under commutators, as $[S_1,S_2]^T = -[S_1,S_2]$ is antisymmetric. This asymmetry between sectors corresponds to the distinction between $\mathbf{16}$ and $\overline{\mathbf{16}}$ spinor representations of SO(10), which transform differently under the gauge group despite having identical internal structure.

\subsection{Representation-Level Correspondence}

The correspondence between nilpotent matrix structure and SO(10) operates at the representation level rather than the gauge algebra level. The matrices reproduce:
\begin{itemize}
\item Three families of 16-plets in both antisymmetric and symmetric sectors
\item Universal (12+4) decomposition in every family
\item Conjugate pairing $\mathbf{16} \leftrightarrow \overline{\mathbf{16}}$ through transpose symmetry
\item $\mathfrak{su}(2)_L \oplus \mathfrak{su}(2)_R$ structure as the span of antisymmetric matrices
\end{itemize}

We emphasize that the nilpotent matrices reproduce the \emph{spinor representation structure} of SO(10), not the 45-dimensional adjoint representation of gauge bosons. The antisymmetric sector generates the chiral gauge algebra $\mathfrak{so}(4,\mathbb{C})$, while the nilpotency constraint $\eta^2=0$ selects special elements that organize as fermion-like multiplets. This suggests that fermionic structure—three generations of matter with (12+4) decomposition—may have an underlying algebraic origin in nilpotent constraints within gauge algebra structures, providing a potential geometric explanation for the Standard Model's generation structure.

\subsection{ Higher Symmetric Nilpotent Sectors: Full 432-Matrix Structure}

In addition to the 48 symmetric zero-diagonal matrices that organize into
three chiral $16$-plets, the enumeration yields 384 further symmetric matrices
satisfying the constraints discussed in Section \ref{sec:matrices} with $c \in \{4,8\}$.  Thus the complete symmetric sector consists of 432
matrices, all of which are nilpotent and normalized in this sense.
These additional 384 matrices exhibit a rich internal structure distinct from
the fermion-like $16$-plets.

\paragraph*{Pauli-basis structure of diagonal blocks:}
Writing each matrix in $2\times 2$ block form
\[
\eta = 
\begin{pmatrix}
A & X \\ X^T & B
\end{pmatrix},
\]
the diagonal blocks admit Pauli-basis decompositions
$A,B \in \mathrm{span}_{\mathbb{C}}\{I, \sigma_x, \sigma_z\}$, with the notable
absence of any $\sigma_y$ component across the entire 432-matrix set.
Within the 48 zero-diagonal $16$-plets, only $\sigma_x$ appears in the diagonal
blocks (when nonzero), as described in Sec.~IV.B.  In contrast, the remaining
384 matrices explore all three directions $\{I,\sigma_x,\sigma_z\}$, subject
to the nilpotency and normalization constraints.

\paragraph*{Decomposition of the 384 additional matrices:}
The 384 symmetric nilpotent matrices beyond the three $16$-plets split into
five algebraically distinct families, determined by which of
$\{I,\sigma_x,\sigma_z\}$ appear in the diagonal blocks $A$ and $B$:
\[
\begin{array}{c|c}
\text{Pauli pattern in }(A,B) & \text{Family size} \\
\hline
I\ \text{only}                & 32 \\
\sigma_z\ \text{only}        & 32 \\
I + \sigma_z                  & 64 \\
I + \sigma_x                  & 64 \\
\sigma_x + \sigma_z           & 192
\end{array}
\]
These five families exhaust the 384 matrices:
\[
32 + 32 + 64 + 64 + 192 = 384.
\]
Each family corresponds to a distinct subset of the Pauli-basis
coefficients $(a_I,a_x,a_z)$ and $(b_I,b_x,b_z)$ in the diagonal blocks
$A = a_I I + a_x\sigma_x + a_z\sigma_z$ and
$B = b_I I + b_x\sigma_x + b_z\sigma_z$.
The systematic exclusion of $\sigma_y$ and the discrete set of allowed
combinations reveal a constrained algebraic variety of symmetric nilpotent
solutions.

The 48 zero-diagonal symmetric matrices singled out in Sec.~II form the three
$\overline{\mathbf{16}}$-like chiral families with a universal $(12+4)$
Pati--Salam decomposition.  The remaining 384 nilpotent matrices differ in two
important respects:
\begin{enumerate}
    \item They do not organize into $16$-plets; their Pauli patterns correspond
    instead to higher-dimensional multiplet structures.
    \item Their diagonal subblocks explore all three directions
    $I$, $\sigma_x$, and $\sigma_z$, in contrast to the purely $\sigma_x$-based
    structure of the fermion-like $16$-plets.
\end{enumerate}
Although we do not attempt a full algebraic classification of these higher
families here, their dimensions $(32,32,64,64,192)$ suggest a constrained
tensor-product-like organization reminiscent of the higher-dimensional adjoint
and tensor representations that appear in continuous grand unified theories.
These higher symmetric nilpotent sectors therefore provide a finite,
discretized shadow of larger gauge-like structures, complementing the chiral
$16$-plet organization of the antisymmetric and symmetric zero-diagonal
subspaces.

\section{Experimental Signatures}

The framework makes concrete, testable predictions through spin-polarized electron scattering experiments.

\textbf{Platform}: Spin-polarized low-energy electron microscopy (SPLEEM)~\cite{Bauer2001} in magnetic thin-film heterostructures provides direct experimental access to the transmission coefficients calculated in Section \ref{sec:scattering}. The experimental setup would be as follows:
\begin{enumerate}
\item Prepare spin-polarized electron beam with controllable incident energy
\item Create potential barriers in engineered thin-film structures
\item Measure energy-dependent spin-resolved transmission: $T_\uparrow(E)$ and $T_\downarrow(E)$
\item Extract spin-flip ratio: $R(E) = T_{\text{flip}}/T_{\text{no-flip}}$
\end{enumerate}


The three antisymmetric families produce distinct, experimentally distinguishable signatures [Fig.~\ref{fig:three_generations}]:

\textbf{Generation 1} (zero diagonal) exhibits the strongest spin-flip behavior, with the reflection coefficient $R(E)$ maximized near threshold. This generation displays a sharp energy onset at threshold $E_{\text{th}}^{(1)}$, followed by a rapid transition to spin-preserving transmission at higher energies. The pronounced threshold behavior and strong spin-flip signature make this generation the most readily observable experimentally.

\textbf{Generation 2} (real diagonal) shows suppressed spin-flip characteristics, with $R(E) < R_{\text{Gen 1}}$ throughout the entire energy range. The threshold behavior exhibits intermediate sharpness at $E_{\text{th}}^{(2)} < E_{\text{th}}^{(1)}$, occurring at lower energy than Generation 1. Additionally, this generation displays partial asymptotic transmission with $T_1 \approx 0.8$, distinguishing it from the complete transmission observed in Generation 1.

\textbf{Generation 3} (complex diagonal) demonstrates the most suppressed spin-flip behavior, with $R(E) \ll R_{\text{Gen 1}}$ across all energies. The threshold occurs at the lowest energy $E_{\text{th}}^{(3)} < E_{\text{th}}^{(2)}$ and exhibits the smoothest onset among the three generations. The asymptotic approach to final transmission values is gradual, reflecting the complex nature of the diagonal blocks and representing the subtlest experimental signature of the three families.

The hierarchy of spin-flip amplitudes $R_1 > R_2 > R_3$ and threshold energies $E_{\text{th}}^{(1)} > E_{\text{th}}^{(2)} > E_{\text{th}}^{(3)}$ provides two independent experimental handles to identify the generation index. 

Materials implementing these coupling structures—whether engineered or naturally occurring in magnetic heterostructures—provide experimental verification or falsification of the framework independent of its theoretical interpretation.

\section{Conclusion}
We have demonstrated that 4×4 nilpotent matrices under quantum mechanical constraints may provide a promising framework to answer some key fundamental unanswered questions such as the existence of three families and the mass hierarchies in the Standard Model. This structure precisely reproduces SO(10) grand unification and its Pati-Salam decomposition: the 2×3×16 organization matches three generations of 16 fermions each, while the internal (12+4) structure—distinguished by the trace of off-diagonal blocks—corresponds to the quark-lepton decomposition across all generations. The emergent $SU(2)_L \times SU(2)_R$ chirality structure arises naturally from nilpotency constraints, with spin-flip behavior distinguishing matrix families experimentally. The correspondence extends to multiple independent structural features, suggesting a non-coincidental connection between discrete nilpotent algebras and continuous gauge symmetries.

The framework makes specific predictions testable in spin-polarized electron scattering experiments. Each generation produces distinct transmission and reflection signatures that can validate or refute the correspondence independent of theoretical interpretation. The predicted experimental variation of the three families provides a concrete test of whether the mathematical structure has physical meaning beyond formal analogy.

Whether this represents a fundamental principle—that gauge symmetries emerge as classical limits of finite algebraic structures—or an elaborate numerical coincidence remains to be determined. The mathematical correspondence appears exact, the predictions are testable, and the implications for understanding the origin of particle multiplicity structure warrant serious investigation.

\section{Acknowledgments}
We would like to thank Abdul Rehman and Fariha Nasir for useful discussions.
\vspace{.5cm}

The code for the analysis in the paper is available on github: \url{https://github.com/dradeelajaib/extensive_nilpotent_search}

\vspace{.5cm}
\appendix
\renewcommand{\theequation}{A\arabic{equation}}
\setcounter{equation}{0}
\section{Transmission Coefficients}
\label{app:transmission}

The energy-dependent transmission coefficients for the three antisymmetric matrix families are given below. We define:
\begin{subequations}
\begin{align}
k &= \sqrt{E^2 - m^2}, \label{eq:k_incoming}\\
k' &= \sqrt{(E - V_0)^2 - m^2}, \label{eq:k_transmitted}
\end{align}
\end{subequations}
representing the incoming and transmitted momenta, respectively.

\subsection*{Generation 1 (Zero Diagonal Blocks)}
The transmission coefficient for Generation 1 is:
\begin{equation}
T_{\text{Gen1}} = \frac{E - V_0}{E} \sqrt{\frac{E^2 - m^2}{(E - V_0)^2 - m^2}} = \frac{E - V_0}{E} \frac{k}{k'}.
\label{eq:TGen1}
\end{equation}

\subsection*{Generation 2 (Real Diagonal Blocks)}
The transmission coefficient for Generation 2 is:
\begin{equation}
T_{\text{Gen2}} = \frac{k k' \left(E(k + k') - k V_0\right)^2}{E(E - V_0)\left(E^2 - m^2 + k k' - E V_0\right)^2}.
\label{eq:TGen2}
\end{equation}

\subsection*{Generation 3 (Complex Diagonal Blocks)}
The transmission coefficient for Generation 3 is:
\begin{equation}
T_{\text{Gen3}} = \frac{(E - V_0) k k' (k + k')^2}{E\left(E^2 - m^2 + k k' - E V_0\right)^2}.
\label{eq:TGen3}
\end{equation}

These expressions are evaluated for electron energies $E > V_0 + m$ to ensure real transmitted momenta. The systematic differences in functional form correspond to the distinct diagonal block structures of each generation: zero, real, and complex entries respectively.

\end{document}